# Fabrication and characterization of femtosecond laser written waveguides in chalcogenide glass

M. Hughes,[a)] W. Yang, and D. Hewak

*Optoelectronics Research Centre, University of Southampton, Southampton SO17 1BJ, United Kingdom*

The authors describe the fabrication of buried waveguides in a highly nonlinear chalcogenide glass, gallium lanthanum sulfide, using focused femtosecond laser pulses. Through optical characterization of the waveguides, they have proposed a formation mechanism and provide comparisons to previous work. Tunneling has been identified as the dominant nonlinear absorption mechanism in the formation of the waveguides. Single mode guidance at 633 nm has been demonstrated. The writing parameters for the minimum propagation loss of 1.47 dB/cm are 0.36 $\mu$J pulse energy and 50 $\mu$m/s scanning speed.

Highly nonlinear glass is an excellent candidate material for optical ultrafast nonlinear devices such as demultiplexers,[1] wavelength converters,[2] and optical Kerr shutters.[3] This is because of its ability to cause nonlinear phase shifts over much shorter interaction lengths than conventional (silica based) devices. Various waveguiding structures such as fibers, proton beam written waveguides, continuous wave laser written waveguides, and femtosecond laser written waveguides could be used to realize such devices. Of these femtosecond laser writing is particularly attractive because as well as having rapid processing times waveguiding structures can be formed below the surface of the glass enabling three-dimensional structures to be fabricated. There have been several studies detailing the fabrication and characterization of waveguides using focused femtosecond laser pulses in phosphate glass,[4] chalcogenide glass,[5] and heavy metal oxide glass.[6] Of these chalcogenide glasses are especially attractive because they have a high nonlinear refractive index and enhanced IR transmission coupled with low maximum phonon energy. Of the chalcogenide glasses gallium lanthanum sulfide (GLS) is probably the most notable with respect to optical nonlinear devices as it has the highest nonlinear figure of merit (FOM) of any glass reported to date,[7] FOM=$n_2/(2\beta\lambda)$, where $\lambda$ is the wavelength, $n_2$ is the real part of the nonlinear refractive index, and $\beta$ is the two-photon absorption coefficient. In this letter we report the fabrication and characterization of buried waveguides written into GLS glass using focused femtosecond laser pulses.

A sample of GLS was prepared by mixing 65% gallium sulfide, 30% lanthanum sulfide, and 5% lanthanum oxide (mol %) in a dry-nitrogen purged glove box. Gallium and lanthanum sulfides were synthesised in-house from high purity gallium and lanthanum precursors in a flowing $H_2S$ gas system; the lanthanum oxide was purchased commercially and used without further purification. The glass was melted in a dry-argon purged furnace at 1150 °C for 24 h before being quenched and annealed at 400 °C for 12 h, it was then cut and polished into a $12 \times 12 \times 5$ mm$^3$ slab.

To write the waveguides a Ti:sapphire laser (Coherent RegA) emitting a train of pulses with a duration of 150 fs, a repetition rate of 250 kHz, and a central wavelength of 800 nm was used. Pulse energy was varied using a variable neutral density filter. The laser beam was focused via a 50× objective [numerical aperture (NA)=0.55] around 200 $\mu$m below the surface of the sample and had a focus spot diameter of around 2 $\mu$m. The sample was mounted on a computer controlled linear motor translation stage which could move in three axes with a resolution of a few nanometers. A series of channels at various pulse energies and translation velocities was written in the sample by translating it perpendicularly to the propagation direction of the laser beam. After processing the end faces of the sample were polished for subsequent characterization.

To obtain guided mode profiles a vertically polarized 633 nm He–Ne laser was coupled into and out of the waveguides with 10× 0.25 NA objectives; the low magnification was needed because of the large mode size of some of the waveguides. The polarization direction was changed with a half wave plate. The near field image was then captured by a charged coupled device camera.

The refractive index change ($\Delta n$) profile was deduced from a quantitative phase image, and taken in the axis the waveguides were written, using quantitative phase microscopy (QPM).[8] The QPM system incorporated a Physik Instrumente nanofocusing Z drive to take in focus and very slightly positively and negatively defocused images; IATIA software was then used to calculate the phase image.

Waveguide loss measurements were taken using the Fabry-Pérot resonance method.[9] By taking into account reflections from its end faces the waveguide structure may be regarded as a resonant cavity. By varying the wavelength of the input light source the output will reach periodic maxima ($I_{max}$) and minima ($I_{min}$), defining $\zeta = I_{max}/I_{min}$. It can be shown[9] that the loss coefficient ($\alpha$) of a waveguide can be calculated using

$$\alpha = -\frac{1}{L} \ln\left(\frac{1}{R}\frac{\sqrt{\xi}-1}{\sqrt{\xi}+1}\right), \quad (1)$$

where $R$ is the reflectivity of the end faces and $L$ is the length of the waveguides. To carry out the experiment a Photonetics Tunics tunable external cavity laser was coupled into and out of the waveguides with 10× 0.25 NA objectives with the output detected by a power meter. The external cavity laser was scanned from 1550 to 1550.5 nm in steps of 0.001 nm and the waveguide output power plotted as a function of input wavelength. Error bounds were calculated from the variance of $\zeta$ and the accuracy to which $R$ was known.

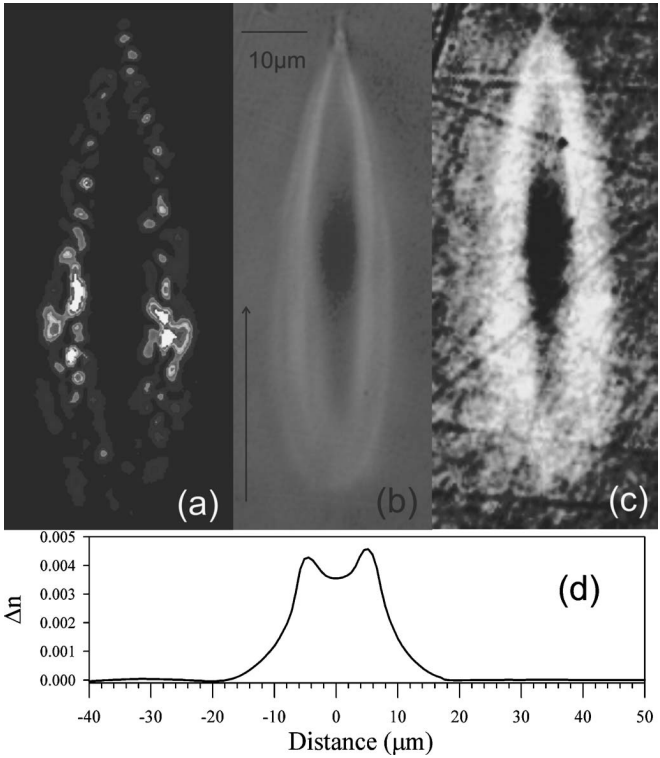

FIG. 1. Guided mode ($\lambda=633$ nm) of waveguide written with a pulse energy of 0.4 $\mu$J and a speed of 50 $\mu$m/s; (b) transmission and (c) reflection optical micrograph of the waveguide end face; (d) refractive index change profile. The arrow shows the propagation direction of the laser used to write the waveguide; all parts of the figure are on the same scale.

Figure 1 shows the guided mode, transmission and reflection optical micrograph of a waveguide written with a pulse energy of 0.4 $\mu$J, and a scan speed of 50 $\mu$m/s. The first thing to note is that the waveguide cross section is highly asymmetric; this can be explained as follows: perpendicular to the propagation direction of the laser the waveguide dimension is given approximately by the beam focal diameter $2\omega_0$, while along the propagation direction it is given by the confocal parameter $b=2\pi\omega_0^2/\lambda$. This results in a large difference in the waveguide sizes in the two directions.[10] Secondly in the transmission optical micrograph a voidlike structure can be seen in the central region of the waveguide, with the waveguide structure surrounding this void. In agreement with this the guided mode shows that the voidlike structure does not actively guide light and it is only the surrounding structure that guides light. Rotating the polarization from vertical ($E$ field along major axis of waveguide) to horizontal increased transmitted power by ~10%. In the reflection optical micrograph the waveguide structure has a high reflectivity indicating it has undergone a positive refractive index change, whereas the voidlike structure has a lower reflectivity indicating that it has undergone a lower refractive index change. The $\Delta n$ profile shows that the voidlike structure has undergone a lower refractive index change than the waveguide structure and that the maximum index change is ~$4.5\times10^{-3}$. These observations indicate that the waveguide structure is formed by a similar mechanism that occurs in phosphate glass;[4] here the femtosecond laser beam induces a modified region in its focal volume that has a lower density and refractive index than the initial glass.

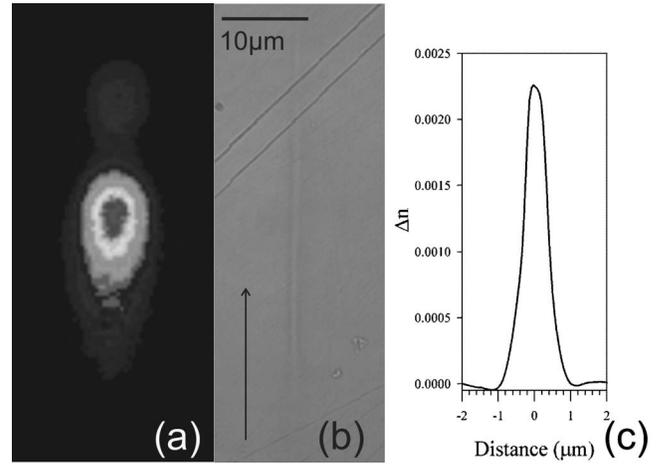

FIG. 2. Guided mode ($\lambda=633$ nm) of waveguide written with a pulse energy of 0.21 $\mu$J and a speed of 200 $\mu$m/s; (b) transmission optical micrograph of the waveguide end face; (c) refractive index change profile. The arrow shows the propagation direction of the laser used to write the waveguide.

Therefore we propose that the voidlike structure was formed by exposure to the focused femtosecond laser beam. The waveguide structure that actively guides light was formed by movement of glass from the femtosecond laser exposed region resulting in a region of higher density and higher refractive index. Laser induced compaction of GLS glass has been shown to result in a positive index change.[11] Interactions between the exposed and densified regions after exposure could result in the voidlike structure having a positive $\Delta n$. Electron dispersive x-ray measurements of the voidlike structure, the waveguide structure, and the surrounding glass found no compositional variation between these regions greater than the system detection limit of around 1%. Micro-Raman measurements of the same regions in the inset of Fig. 3 show no significant variations. The Raman spectra of GLS consist of two broad bands located at 150 and 340 cm$^{-1}$ (Ref. 12) making structural variations difficult to distinguish. We therefore propose that any structural modifications that occur are subtle, such as a bond angle change.

Figure 2 shows the guided mode at 633 nm and transmission optical micrograph of a waveguide written with a pulse energy of 0.21 $\mu$J and a scan speed of 200 $\mu$m/s. A Gaussian profile can be fitted to the guided mode with a correlation coefficient of 0.986 in the horizontal and 0.923 in the vertical, indicating the waveguide is single mode at 633 nm. In the transmission optical micrograph a voidlike structure is not clearly visible as in Fig. 1. The $\Delta n$ profile also shows no voidlike structure and that the maximum index change has fallen to ~$2.3\times10^{-3}$. The absence of a voidlike structure may be because it is too small to be resolved or that a different formation mechanism occurs at this pulse energy. The linear shape of the waveguide indicates that filamentation caused by self-focusing may have occurred in the formation of this waveguide.

Several mechanisms for nonlinear absorption of femtosecond pulses have been considered in the literature; the principal ones are multiphoton ionization (MPI), tunneling, and avalanche ionization. It has been proposed by several authors that avalanche ionization is of little importance in the absorption of subpicosecond pulses in transparent materials.[13–15] We therefore propose that avalanche ioniza-tion is not significant in the formation of these waveguides. In order to determine whether MPI or tunneling dominate absorption the Keldysh parameter ($\gamma$) can be calculated using.[14,16,17]

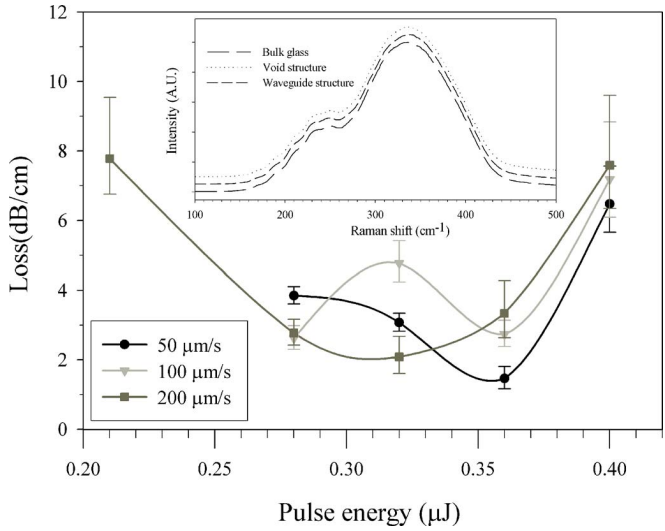

FIG. 3. Waveguide losses as a function of writing pulse energy at various scanning velocities. Lines are a guide for the eye. Inset shows micro-Raman measurements of the bulk glass, the voidlike structure, and the waveguide structure of a waveguide written with pulse energy of 1.74 µJ.

$$\gamma = \frac{\Omega \sqrt{m_{\text{red}}\Delta}}{eE}, \qquad (2)$$

where $\Omega$ is the laser frequency, $m_{\text{red}}$ is the reduced effective mass, $\Delta$ is the band gap energy which is 2.48 eV for GLS, $e$ is the electron charge, and $E$ is the peak electric field intensity. The reduced effective mass is related to the effective mass in the conduction band ($m_c$) and valence band ($m_v$), which are both assumed to be equal to the free electron mass, by $1/m_{\text{red}} = 1/m_c + 1/m_v$. For $\gamma > 1$ MPI is dominant, for $\gamma < 1$ tunneling is dominant. The waveguides investigated in this study were written with pulse energies of 0.21–1.74 µJ, which correspond to a Keldysh parameter ($\gamma$) of 0.11–0.04; this indicates that tunneling was the dominant nonlinear absorption process in the formation of all the waveguides investigated in this study.

Figure 3 shows the loss measurements of the waveguides as a function of pulse energy at various scanning velocities. For waveguides written at 200 and 50 µm/s, an optimum pulse energy reached at 0.32 and 0.36 µJ/pulse giving losses of 2.08 and 1.47 dB/cm, respectively. However, for waveguides written at 100 µm/s there is no apparent optimum writing pulse energy. This may have been because one of the waveguides was damaged or written through a flaw or crystal in the glass. The Fabry-Pérot scans showed regular maxima and minima with no indication of higher order modes present. We propose that the optimum writing parameters occur because at higher pulse energies damage occurs to the glass which increases loss and at lower pulse energies the induced refractive index change is not large enough to fully confine the beam. The minimum reported loss in this work of 1.47 dB/cm compares to minimum losses in femtosecond laser written waveguides of 1.0 dB/cm in aluminosilicate glass,[18] 0.8 dB/cm in fused silica,[19] 0.25 dB/cm in phosphate glass,[10] and 0.8 dB/cm in heavy metal oxides glass.[6] We believe the loss reported in this work could be improved by further optimization of pulse energy and scan speed, using double pulse femtosecond lasers,[19] astigmatically shaping the writing beam to reduce the asymmetry of the waveguide cross section[10] and annealing to reduce any internal stresses in the glass.

In summary we have proposed a formation mechanism for femtosecond laser written waveguides in GLS glass based on optical characterization and comparisons to previous work. Tunneling has been identified as the nonlinear absorption mechanism in the formation of the waveguides by calculation of the Keldysh parameter. Single mode operation at 633 nm has been demonstrated. The writing parameters for the minimum achieved loss of 1.47 dB/cm are 0.36 µJ pulse energy and 50 µm/s scanning speed. We believe these waveguides show promise for all optical device applications in an optical chip configuration.